\documentclass{article}
\usepackage{spconf,amsmath,graphicx}
\usepackage{algpseudocode}
\usepackage{amsfonts} 
\usepackage{algorithm} 
\usepackage{caption}
\usepackage[caption=false,font=footnotesize]{subfig}
\usepackage{xcolor}
\title{Enhanced Channel Estimation in mm-Wave MIMO Systems Leveraging Integrated Communication and Sensing}
\name{Silvia Mura$^\dagger$, Marouan Mizmizi$^\dagger$, Umberto Spagnolini$^\dagger$ and Athina Petropulu$^*$}
\address{$^\dagger$Dipartimento di Elettronica, Informazione e Bioingegneria,
Politecnico di Milano\\
$^*$Dept. of Electrical and Computer Engineering, Rutgers University}

\begin{document}

\maketitle
\begin{abstract}
This paper tackles the challenge of wideband MIMO channel estimation within indoor millimeter-wave scenarios. Our proposed approach exploits the integrated sensing and communication paradigm, where sensing information aids in channel estimation. The key innovation consists of employing both spatial and temporal sensing modes to significantly reduce the number of required training pilots. Moreover, our algorithm addresses and corrects potential mismatches between sensing and communication modes, which can arise from differing sensing and communication propagation paths. Extensive simulations demonstrate that the proposed method requires $4\times$ less pilots compared to the current state-of-the-art, marking a substantial advancement in channel estimation efficiency.
\end{abstract}
\section{Introduction}
\label{sect:introduction}

Future wireless networks are exploiting higher frequencies, notably millimeter waves (mmWaves), to meet the ever-growing demand for user throughput. However, mmWave propagation poses significant challenges due to high path and blockage attenuation, particularly for indoor scenarios~\cite{rappaport2002wireless}. In this context, the channel state information (CSI) acquisition assumes pivotal significance for accurate signal decoding. A prevalent approach to CSI estimation consists of multiplexing known pilots with data, thereby enabling CSI acquisition at the receiver's side~\cite{alkhateeb2014channel}. The effectiveness of CSI acquisition depends on the estimation method and the number of pilots.

Numerous channel estimation techniques have been proposed spanning conventional Bayesian methodologies, e.g., least square (LS), maximum likelihood, and minimum mean square error (MMSE), along with novel approaches such as compressed sensing (CS)~\cite{8322235}, low-rank~\cite{mizmizi2021channel}, and machine learning techniques~\cite{8752012}.
CS reduces pilot overhead by exploiting channel sparsity, making it a viable choice for mmWave communications, which exhibit a limited number of dominant scatterers. However, in indoor settings, CS-based estimation can be computationally intensive due to the increased number of scatterers, which leads to rapid CSI variations and requires higher training samples compared to outdoor scenarios~\cite{chen2021off}. 

Integrated Sensing and Communication (ISAC) is a novel approach that combines sensing and communication at the transmitter~\cite{liu2020joint, zhang2021overview}. ISAC can leverage the correlation between sensing and communication channels to reduce the number of training samples required for channel estimation. Prior research has predominantly focused on examining this correlation in outdoor MIMO ISAC systems. For instance,~\cite{ali2020passive} exploits the sensing covariance to estimate communication channel covariance, reducing beam training overhead.
By leveraging the correlation between the uplink channel and sensing,~\cite{9990573} suggests a sensing-aided Kalman filter-based method to enhance CSI estimation accuracy.~\cite{jiang2022sensing} formulates orthogonal time frequency space channel estimation as a sparse recovery problem by utilizing the sensing information to determine the delay and Doppler support.~\cite{9805471} proposes a turbo sparse Bayesian inference for target detection and channel estimation in narrowband ISAC systems. 

\begin{figure}[b!]
\vspace{-0.5cm}
    \centering
    \includegraphics[width=0.60\columnwidth]{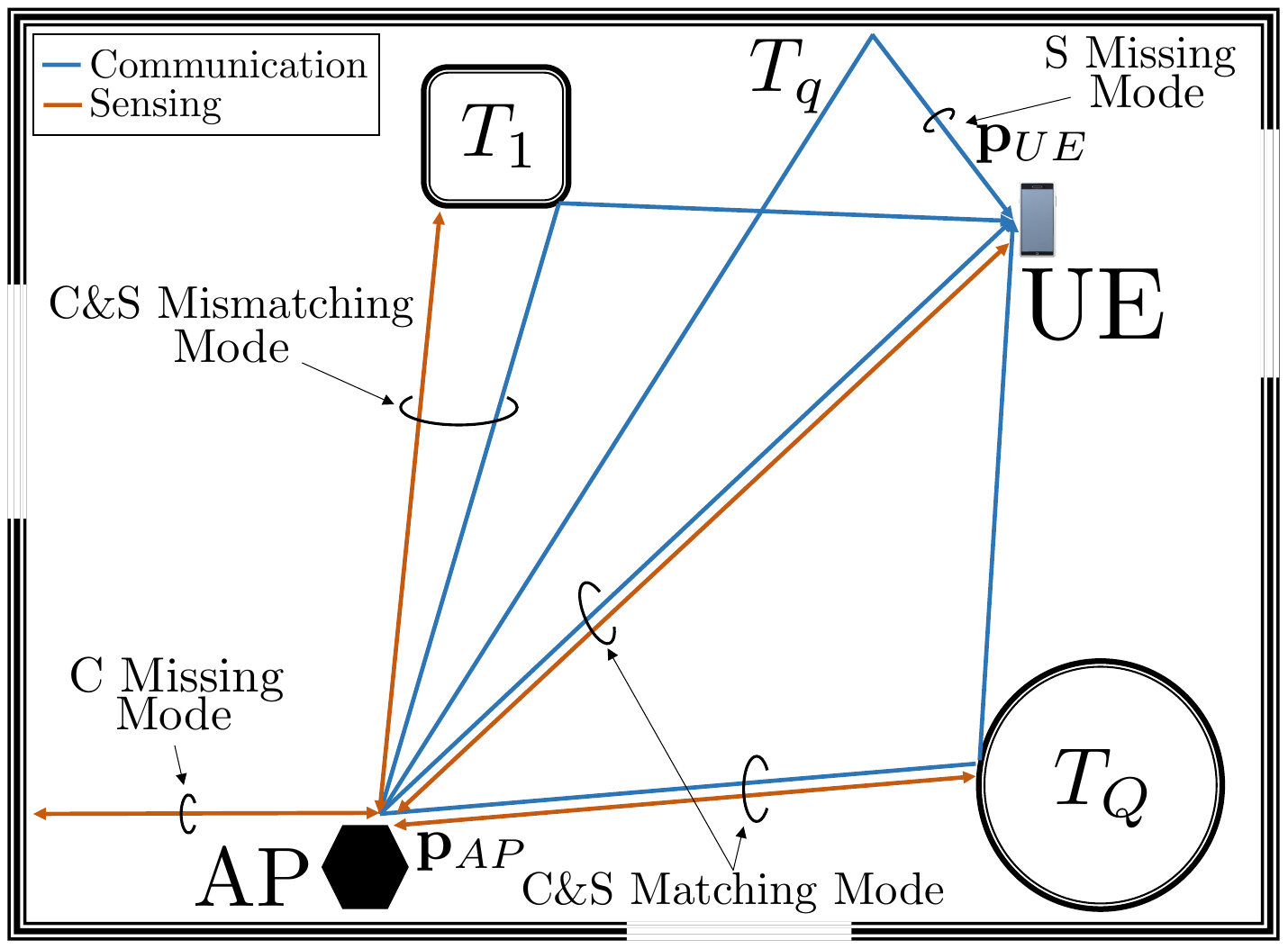}
    \caption{Possible mismatches between communication and sensing channel modes in the reference indoor scenario.}
    \label{fig:ReferenceScen}
\end{figure}

In ISAC systems, the location of scattering objects is estimated and then used to determine the space-time modes of the communication channel, i.e. the delay and angular characteristics of the channel. However, the sensing operation is typically done in a monostatic configuration, while communication channels use a bistatic setup. This difference in setup, as well as the presence of extended scatterers, may lead to communication and sensing (C\&S) mode mismatch. As depicted in Fig.~\ref{fig:ReferenceScen}, C\&S mismatch may involve discrepancies between sensing and communication space-time modes~\cite{10008630}, as well as missing modes in the sensing channel, and vice versa. Most previous works in \cite{ali2020passive}-\cite{9805471} have not explored this aspect, which may lead to inaccurate channel estimation and system performance degradation.

This paper presents a sensing-aided CS channel estimation algorithm that, differently from prior works, addresses C\&S mode mismatch by defining the initial communication space-time modes based on the sensing signals and compensating for the possible C\&S mismatches. Subsequently, the communication channel modes are augmented by identifying communication missing modes, related to scatterers not visible in the sensing signals, within an angular and temporal mode codebook, tailored to the indoor scenario. This approach is suitable for wideband channel estimation and represents a major improvement over existing methods.
Numerical simulations demonstrate that the sensing-assisted channel estimation method requires 4$\times$ less of training pilots compared to current state-of-the-art solutions.

The paper is organized as follows: Section II defines the system and channel model, Section III details the proposed sensing-assisted CS algorithm, numerical results are discussed in Section IV, and Section V draws the conclusions.

\vspace{-0.2cm}
\section{System and Channel Model}\label{sect:system_model}

\begin{figure}[b]
    \vspace{-0.7cm}
    \centering
    \includegraphics[width=0.85\columnwidth]{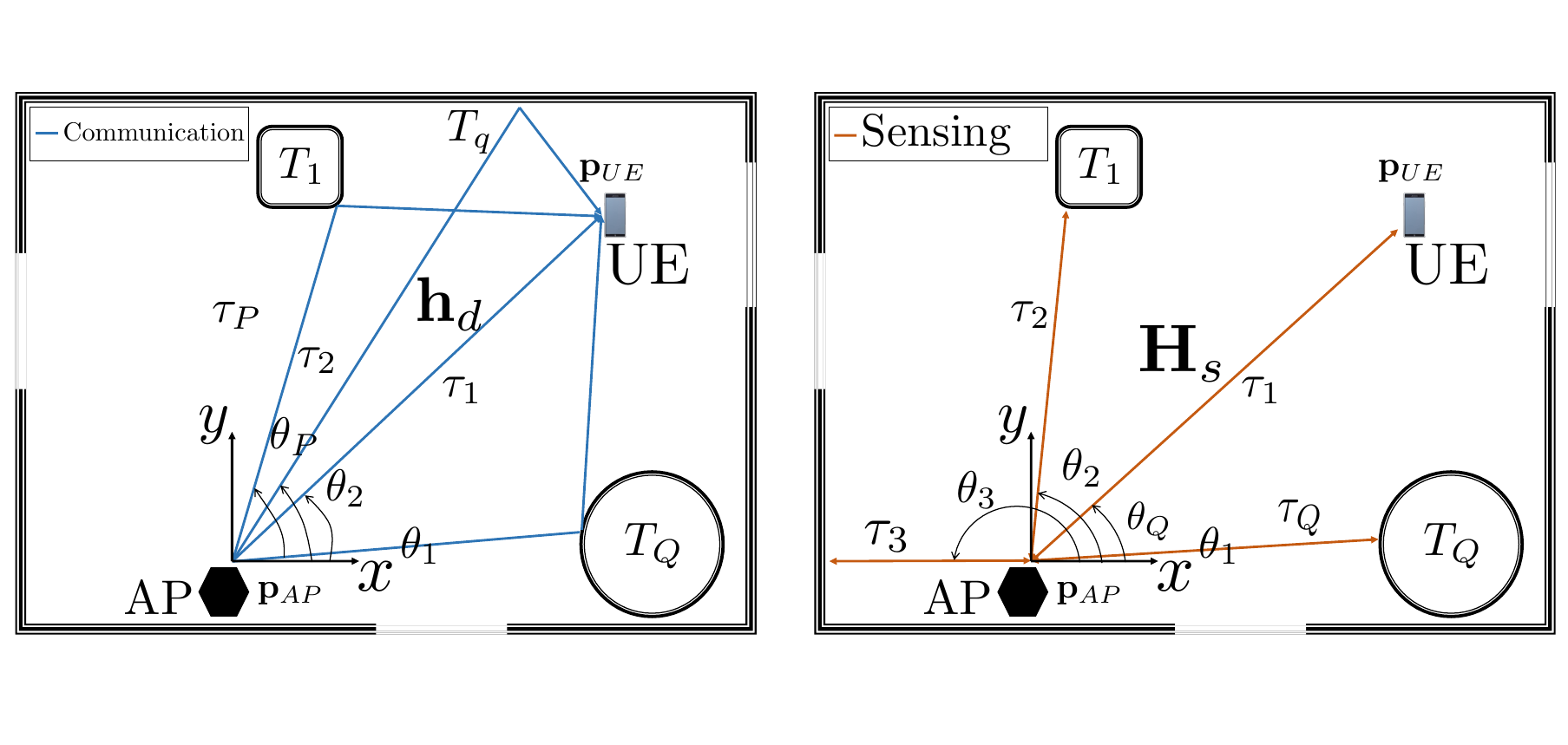}  
    \vspace{-0.5cm}
    \caption{Reference indoor scenario}
    \label{fig:IndoorISAC}
\end{figure}

Consider the indoor setting depicted in Figure~\ref{fig:IndoorISAC}. Here, the access point (AP) includes two $N$ elements ULA antenna arrays for simultaneous transmission and reception to facilitate ISAC functionality. Within each channel coherence interval, the AP communicates with a single antenna user equipment (UE) in downlink mode, while concurrently sensing $Q$ extended targets, including the UE, namely $T_1, ..., T_Q$ in Fig.~\ref{fig:IndoorISAC}. Subsequently, the UE transmits its data to the AP in uplink mode. 
The transmitted signal is an orthogonal frequency-division waveform with a bandwidth $B = K \Delta f$, where $K$ and $\Delta f$ denote the number of subcarriers and the subcarrier spacing, respectively. Within the $k$th sub-carrier, the downlink signal transmitted by the AP is expressed as
\begin{align}
    \mathbf{x} [k] = \mathbf{f} [k] s_d [k],
\end{align}
where $s_d[k]$ denotes the $k$th transmitted symbol such that $\mathbb{E}[s_d[k] s_d [m]^*]$ = $\sigma_s^2 \delta[k-m]$, with power $\sigma_s^2$, and $\mathbf{f}[k] \in \mathbb{C}^{N \times 1}$ represents the precoding vector at the $k$th subcarrier. The sensing signal received by the AP is
\begin{align}\label{eq:RxsignalBS}
    \mathbf{r}[k] = \mathbf{H}_s [k]\mathbf{x} [k] + \mathbf{n}_s[k],
\end{align}
where $\mathbf{H}_s [k] \in \mathbb{C}^{N \times N}$ denotes the sensing channel matrix and $\mathbf{n}_s[k] \sim\mathcal{CN}(0, \sigma^2_s\, \mathbf{I}_{N})$ is the noise.

The downlink signal received by the UE is expressed as
\begin{align}
    y_d [k] = \sqrt{\rho_d} \mathbf{h}_d[k] \mathbf{x} [k]+ n_d[k],
\end{align}
where $\rho_d$ represents the average downlink received power, $\mathbf{h}_d[k] \in \mathbb{C}^{1 \times N}$ denotes the communication channel vector such that $\mathbb{E}[\mathbf{h}_d\mathbf{h}_d^\mathrm{H}] = N$ and $n_d[k] \sim\mathcal{CN}(0, \sigma^2_n)$ is the additive noise. 

In the uplink operation, the received signal is expressed as
\begin{align}
    \mathbf{y}_u[k] = \sqrt{\rho_u} \mathbf{h}_u [k] s_u[k] + \mathbf{n}_u[k],
\end{align}
where  $\rho_u$ is the average uplink received power, $\mathbf{h}_u[k] \in \mathbb{C}^{N \times 1}$ denotes the uplink communication channel vector such that $\mathbb{E}[\mathbf{h}_u^\mathrm{H}\mathbf{h}_u] = N$, and $\mathbf{n}_u [k] \sim\mathcal{CN}(0, \sigma^2_n\,\mathbf{I}_N)$ is the additive noise. The symbol transmitted by the UE on the $k$th sub-carrier, denoted as $s_u[k]$, is designed such that $\mathbb{E}[s_u[k]s_u^*[m]] = \sigma_s^2 \delta[k-m]$. Here, $\sigma_s^2$ refers to the transmitted power. Among the transmitted symbols $\mathbf{s}_u = [s_u[0], s_u[1], \ldots, s_u[K-1]]$, there are $K_p < K$ pilots regularly placed across the subcarriers, explicitly for channel estimation purposes.

\subsection{Communication Channel Model}

The high free-space pathloss that is a characteristic of mmWave propagation leads to limited space-time selectivity. For this reason, we adopt a block-fading clustered channel representation, based on the extended Saleh-Valenzuela model, which allows us to accurately capture the mathematical structure present in mmWave channels \cite{4022660}. Hence, the uplink channel in the frequency domain can be expressed as
\begin{align}\label{eq:comChannel}
\begin{split}
    \mathbf{h}_u[k] = & \sqrt{\frac{N}{P}} \sum_{p=1}^P \alpha_p \mathbf{a}(\theta_p) e^{-j\frac{2 \pi k \tau_p}{K}},
    \end{split}
\end{align}
where $P$ denotes the number of propagation paths, $\alpha_p$ is the complex gain, $\mathbf{a}(\theta)$ denotes the AP array response vector, $\theta_p$ and $\tau_p$ denote the angle of arrival, and delay of the $p$th path, respectively. Herein, we assume channel reciprocity \cite{7533487}, hence, $\mathbf{h}_d[k] = \mathbf{h}_u^\mathrm{T}[k]$.

\subsection{Sensing Channel Model}

Similarly, the sensing channel is expressed as
\begin{align}\label{eq:SensChannel}
    \mathbf{H}_s[k] = \sum_{q=1}^{Q} \beta_q \mathbf{a}(\theta_{q}) \mathbf{a}(\theta_{q})^\mathrm{H} G[k] e^{-j\frac{2\pi k \tau_q}{K}},
\end{align}
where $\theta_q$ is the azimuth angle related to the $q$th target, $\tau_{q}= 2 d_q/c$ denotes the two-way propagation delay between the AP and the $q$th target, with $d_q$ being the distance between the AP and the $q$th target. The model of the scattering coefficient in \eqref{eq:SensChannel} follows the radar equation \cite{skolnik1980introduction}
\begin{align}\label{eq:beta}
    \beta_q = \sqrt{\frac{\lambda^2 N^2}{(4 \pi d_q)^4} \Gamma_{q}} \; e^{j\xi_q},
\end{align}
where $\Gamma_q$ denotes the radar cross-section of the target and $\xi_q$ is an additional random phase term accounting for the Tx/Rx circuitry and Doppler shift due to the target's mobility.

\vspace{-0.2cm}
\section{Sensing-aided Channel Estimation}

This section details the proposed sensing-aided channel estimation method. 
Let us reformulate the communication channel vector in \eqref{eq:comChannel} as
\begin{equation}\label{eq:comChannel2}
    \mathbf{h}_u [k] = \mathbf{A}(\boldsymbol{\theta}) \mathbf{T}[\boldsymbol{\tau}, k] \boldsymbol{\alpha} =  \boldsymbol{\Phi}[k] \boldsymbol{\alpha}
\end{equation}
where $\mathbf{A}(\boldsymbol{\theta}) = \sqrt{\frac{N}{P}}[\mathbf{a}(\theta_1),\ldots,\mathbf{a}(\theta_P)] \in \mathbb{C}^{N\times P}$, $\mathbf{T}[\boldsymbol{\tau}, k] = \mathrm{diag}\left(e^{-j\frac{2\pi k \tau_{1}}{K}},\ldots, e^{-j\frac{2\pi k \tau_{P}}{K}}\right) \in \mathbb{C}^{P \times P}$  and $\boldsymbol{\alpha} \in \mathbb{C}^{P \times 1}$ denotes the communication channel coefficient vector. The matrix $\boldsymbol{\Phi}[k]$ represents the combined space-time modes of the communication channel.
The goal of the proposed method is to estimate the matrix $\boldsymbol{\Phi}[k]$ and the channel coefficient vector $\boldsymbol{\alpha}$ by exploiting sensing information with limited pilots resources.
 
\subsection{From Sensing to Communication Modes}

\begin{figure}[b]
    \vspace{-0.5cm}
    \centering
    \includegraphics[width=0.45\columnwidth]{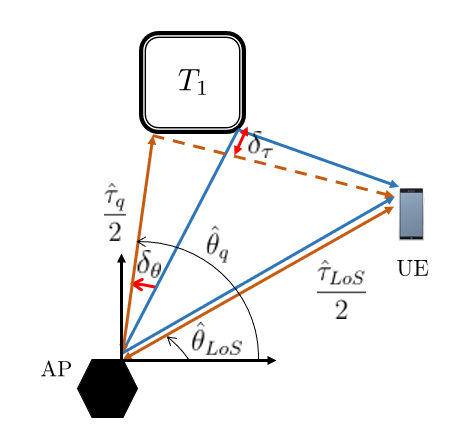}
    \caption{Initial Communication mode estimate and Communication and Sensing mismatch}
    \label{fig:mismatch}
\end{figure}

The estimated sensing space-time modes of the targets, specifically  $\hat{\boldsymbol{\tau}}$ and $\hat{\boldsymbol{\theta}}$, are derived from the signal in \eqref{eq:RxsignalBS} using a range-angle compression as in \cite{8059539}. To distinguish between line-of-sight (LoS) and non-line-of-sight (NLoS) modes, we employ the initial access procedure detailed in \cite{mizmizi2022fastening}. This results in the decomposition of $\hat{\boldsymbol{\tau}}$ into $\hat{\tau}_{LoS}$ and a set of NLoS delays denoted as $\hat{\tau}_1, \hat{\tau}_2, \ldots, \hat{\tau}_{Q-1}$, as well as the partitioning of $\hat{\boldsymbol{\theta}}$ into $\hat{\theta}_{LoS}$ and the NLoS angles denoted as $\hat{\theta}_1, \hat{\theta}_2, \ldots, \hat{\theta}_{Q-1}$.

Assuming only single reflections, the temporal modes are defined by employing the geometric cosine law, such as
\begin{align}\label{eq:tau}
    \tilde{\tau}_q = \frac{\hat{\tau}_q }{2}+ \sqrt{\frac{\hat{\tau}_{LoS}^2}{4}  +\frac{\hat{\tau}_{q}^2}{4}-\frac{\hat{\tau}_{LoS} \, \hat{\tau}_q}{2} \, \cos(\hat{\theta}_q - \hat{\theta}_{LoS})},
\end{align}
and $\tilde{\tau}_{LoS} = \hat{\tau}_{LoS} / 2$.
To account for C\&S mismatch, we assume that the space-time features are affected by an additional error, such that: $\bar{\theta}_q = \hat{\theta}_q + \delta_\theta$ and $\bar{\tau}_q = \tilde{\tau}_q + \delta_\tau$, where $\delta_\theta$ and $\delta_\tau$ represent the additional error arising from C\&S mismatch, as illustrated in Fig. \ref{fig:mismatch}. These additional errors $\boldsymbol{\delta} = [\boldsymbol{\delta}_{{\theta}}\,^{\mathrm{T}}, \boldsymbol{\delta}_{{\tau}}\,^{\mathrm{T}}]^{\mathrm{T}} \in \mathbb{R}^{2Q \times 1}$ will be estimated in the subsequent analysis. Using $\bar{\boldsymbol{\theta}}$ and $\bar{\boldsymbol{\tau}}$, we can compute the communication space-time modes $\hat{\boldsymbol{\Phi}}_{\boldsymbol{\delta}}[k]$ as in \eqref{eq:comChannel2}, which will be further refined in the following.

\subsection{Modal Mismatch Compensation and Channel Estimation}

Estimation of the channel parameters and compensation of mismatch can be formally framed as the following optimization problem:
\begin{alignat}{2}
    &\underset{\boldsymbol{\alpha}, \boldsymbol{\delta}}{\mathrm{minimize}} && \left\|\bar{\mathbf{y}} -\bar{\mathbf{\Phi}}_{\boldsymbol{\delta}}\hat{\boldsymbol{\alpha}} \odot \bar{\mathbf{s}}\right\|_2 + \lambda \left\|\hat{\boldsymbol{\alpha}}\right\|_1 \label{eq:obj},
\end{alignat}
where $\bar{\mathbf{y}}$ = $[\mathbf{y}[0]^\mathrm{T}, \ldots,\mathbf{y}[K_p-1]^\mathrm{T}]^\mathrm{T}$ represents the received pilot symbols, $\bar{\mathbf{\Phi}}_{\boldsymbol{\delta}}$ =  $[\hat{\mathbf{\Phi}}_{\boldsymbol{\delta}}[1]^\mathrm{T}, \ldots,\hat{\mathbf{\Phi}}_{\boldsymbol{\delta}} [K_p]^\mathrm{T}]^\mathrm{T} \in \mathbb{C}^{NK_p \times Q}$
is the estimated modes matrix at the pilot subcarriers, 
$\hat{\boldsymbol{\alpha}} \in \mathbb{C}^{Q \times 1}$
denotes the vector of channel coefficients to be estimated,
$\bar{\mathbf{s}} = \bar{\mathbf{s}}_u \otimes \mathbf{1}_N \in \mathbb{C}^{NK_p \times 1}$ 
represents the transmitted pilot symbols where $\bar{\mathbf{s}}_u = [s_u[0], ...,s_u[K_p-1]]^\mathrm{T}$ and $\lambda > 0$ serves as a hyperparameter. 

\begin{algorithm}[!t] 
\caption{Sensing-Aided Channel Estimation}\label{alg:TotAlg}
\begin{small}
\begin{algorithmic}

    \Statex{\textbf{Input}:} $\bar{\mathbf{y}}$, $\bar{\mathbf{s}}$, $\bar{\mathbf{\Phi}}$, $\hat{\theta}_{LoS}$, $\hat{\tau}_{LoS}$, $\varepsilon$
    
    \Statex{\textbf{Initialization}: i = 0, $\boldsymbol{\delta}^{(i)} = \mathbf{0}_{2Q \times 1}$, $\hat{\boldsymbol{\alpha}}^{(i)} = \mathbf{0}_{Q \times 1}$,}
    
    \Statex{\textbf{Step 1}:} Channel Coefficient Estimation
    \Statex{\hspace{0.6cm} $\hat{\boldsymbol{\alpha}}^{(i+1)} = \underset{\boldsymbol{\alpha}}{\mathrm{minimize}} \left\|\bar{\mathbf{y}} -\bar{\mathbf{\Phi}}^{(i)}_{\boldsymbol{\delta}}\hat{\boldsymbol{\alpha}}^{(i)} \odot \bar{\mathbf{s}}\right\|_2+\lambda\left\|\hat{\boldsymbol{\alpha}}^{(i)}\right\|_1$}
    
    \Statex{\textbf{Step 2}:} Mode Mismatch Compensation
    
    \Statex{\hspace{0.6cm} $\boldsymbol{\delta}^{(i+1)} = \underset{\boldsymbol{\delta}}{\mathrm{minimize}} \left\|\bar{\mathbf{y}} - {\bar{\mathbf{\Phi}}}^{(i)}_{\boldsymbol{\delta}}\hat{\boldsymbol{\alpha}}^{(i+1)} \odot \bar{\mathbf{s}}\right\|_2$}

    \Statex{\hspace{0.6cm} Update $\bar{\mathbf{\Phi}}^{(i+1)}_{\boldsymbol{\delta}}$ with $\boldsymbol{\delta}^{(i+1)}$.}
    
    \Statex{\textbf{Step 3}:} Mode Augmentation
    
    \Statex{\,\,{\hspace{1cm}$\boldsymbol{\epsilon}^{(i+1)} = \bar{\mathbf{y}} - {\bar{\mathbf{\Phi}}}^{(i+1)}_{\boldsymbol{\delta}}{\hat{\boldsymbol{\alpha}}}^{(i+1)} \odot \bar{\mathbf{s}}$,}}
    
    %\Statex{\,\,\hspace{1cm}$[x_1,x_2] = \underset{x}{\mathrm{max}}\,\,||\,(\,[{\bar{\mathbf{\Phi}}}^{(i+1)}_{\boldsymbol{\delta}}]_{x}\,)^\mathrm{H}\boldsymbol{\epsilon}^{(i+1)}\,||^2$},

    \Statex{\,\,\hspace{1cm}${\ell} = \mathrm{argmax}\,\,||\,(\,{\bar{\mathbf{\Phi}}}^{(i+1)}_{\boldsymbol{\delta}}\,)^\mathrm{H}\boldsymbol{\epsilon}^{(i+1)}\,||^2$, $\mathbf{b} = [{\bar{\mathbf{\Phi}}}^{(i+1)}_{\boldsymbol{\delta}}]_{\ell}$}
    
    \Statex{\,\,\hspace{1cm}$\tilde{{\ell}}= {\mathrm{argmax}}\,\,||(\bar{\mathbf{\Phi}}_c)^\mathrm{H}\boldsymbol{\epsilon}^{(i+1)}||^2$, $\tilde{\mathbf{b}} = [\bar{\mathbf{\Phi}}_c]_{\tilde{\ell}}$}

    \Statex{\,\,\hspace{1cm}\textbf{if} $||\,\tilde{\mathbf{b}}^\mathrm{H}\boldsymbol{\epsilon}^{(i+1)}\,||^2 > ||\,\mathbf{b}^\mathrm{H}\boldsymbol{\epsilon}^{(i+1)}\,||^2$}

    \Statex{\,\,\hspace{1.4cm}{$\bar{\mathbf{\Phi}}^{(i+1)}_{\boldsymbol{\delta}} = [{\bar{\mathbf{\Phi}}}^{(i+1)}_{\boldsymbol{\delta}} | \tilde{\boldsymbol{b}}]$,}}
    
    \Statex{\,\,\hspace{1.4cm}{$\boldsymbol{\delta}^{(i+1)} = [\boldsymbol{\delta}^{(i+1)} | \mathbf{0}_{2 \times 1}]$}}
    
    \Statex{\,\,\hspace{1cm}\textbf{end}}

    \Statex{\,\,Compute estimated channel: $\hat{\mathbf{h}}_u = \bar{\mathbf{\Phi}}^{(i+1)}_{\boldsymbol{\delta}} \, \hat{\boldsymbol{\alpha}}^{(i+1)}$,}
    
    \Statex{\,\,Terminate if $\frac{||\hat{\boldsymbol{\alpha}}^{(i+1)}-\hat{\boldsymbol{\alpha}}^{(i)}||_2^2}{||\hat{\boldsymbol{\alpha}}^{(i+1)}||_2^2} \leq \varepsilon$, otherwise return to \textbf{Step 1}.}

\end{algorithmic}
\end{small}
\end{algorithm}

The optimization problem in \eqref{eq:obj} is not convex. Hence, we propose the algorithm \ref{alg:TotAlg} that iterates through three pivotal steps: channel coefficient estimation, mode mismatch compensation, and mode augmentation. In the first step, the cost function is minimized with respect to the channel coefficients  $\hat{\boldsymbol{\alpha}}$ through orthogonal matching pursuit (OMP) \cite{wang2012generalized}, while the second step involves estimating C\&S mismatches $\boldsymbol{\delta}$ based on the reconstructed channel. Due to the non-convex nature of the objective function, the optimal mismatch vector $\boldsymbol{\delta}^{(i+1)}$ is determined using genetic algorithms \cite{mirjalili2019genetic}. Before proceeding to the next iteration, the space-time mode matrix $\bar{\mathbf{\Phi}}^{(i)}_{\boldsymbol{\delta}}$ is updated using the optimal mismatch. Additionally, we enhance the space-time mode matrix by adding a new mode from the codebook space-time matrix $\bar{\mathbf{\Phi}}_c =[{\mathbf{\Phi}}_c[1]^\mathrm{T}, \ldots,{\mathbf{\Phi}}_c [K_p]^\mathrm{T}]^\mathrm{T}$ that best aligns with the estimation error $\boldsymbol{\epsilon}^{(i+1)}$. The codebook is determined according to the room geometry, represented by $\boldsymbol{\theta}_c$ and $\boldsymbol{\tau}_c$, as $\mathbf{\Phi}_c[k] = \mathbf{A}(\boldsymbol{\theta}_c) \mathbf{T}[\boldsymbol{\tau}_c, k]$. A novel space-time mode $\tilde{\mathbf{b}}$, with the highest correlation to the estimation error $\boldsymbol{\epsilon}^{(i+1)}$, is obtained and, if more correlated to the residual than the initial mode, it is incorporated into the current space-time mode matrix $\bar{\mathbf{\Phi}}^{(i+1)}_{\boldsymbol{\delta}}$. Finally, the uplink channel over all the subcarriers is obtained by interpolating the estimated channel ${\hat{\mathbf{h}}_u}$. 

\section{Numerical Results}\label{sect:numerical_results}

Herein, we consider an indoor office scenario. The carrier frequency is $60$ GHz and both sensing and communication channels are simulated using Matlab Ray-Tracing package. We assume Ricean fading with a 4 dB Ricean factor and $N=8$ antenna elements arrays~\cite{7533487}. The performance is evaluated in terms of mean square error (MSE), i.e.,
\begin{align}
    \mathrm{MSE} = \mathbb{E}_k \left[\left\| \hat{\mathbf{h}}_u[k]- {\mathbf{h}}_u[k]\right\|^2\right]
\end{align}
and symbol error rate (SER), defined as the ratio between the erroneous estimated symbols and the total number of data symbols. The estimated data symbols at the AP are
\begin{align}
    \hat{{s}}_u [k] = \textbf{w}_u^{\mathrm{H}}[k]\textbf{y}_u [k],
\end{align}
where $\textbf{w}_u[k]$ denotes the MMSE combiner, derived as in \cite{mizmizi2021channel}, using the estimated channel $\hat{\mathbf{h}}_u[k]$.
The performance is evaluated by varying the percentage of pilots, namely $\eta = K_p/K = 5 \%$ and $20\%$ and the communication SNR at the antenna, defined as $\gamma_0 = \sigma_s^2\rho_u/\sigma_n^2$. 

\begin{figure}[t!]
    \centering   \includegraphics[width=0.85\columnwidth]{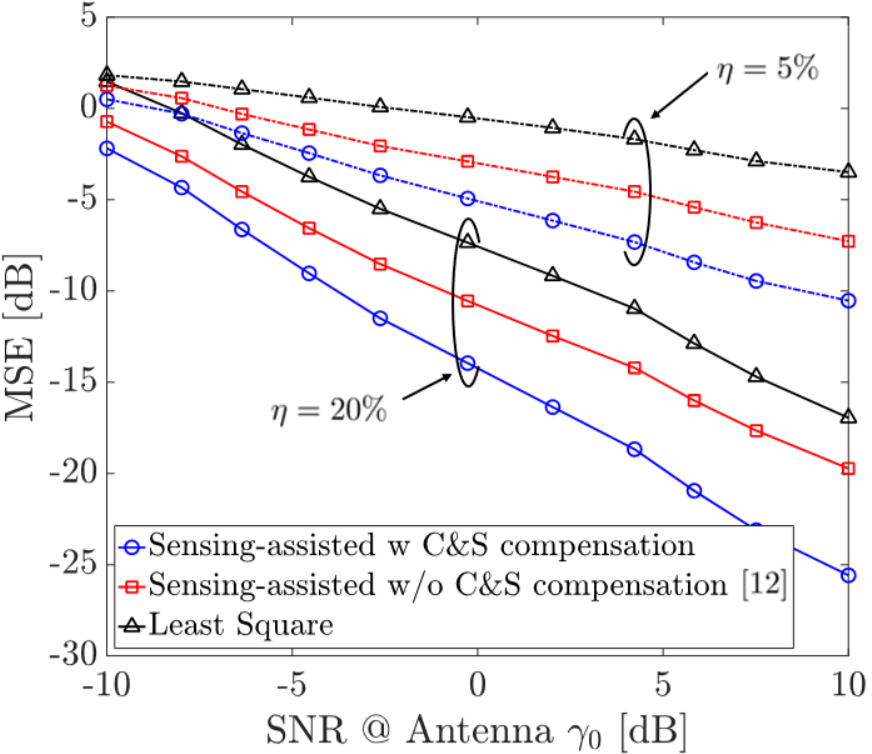}
    \caption{MSE vs Communication SNR at the antenna}
    \label{fig:mse}
    \vspace{-0.3cm}
\end{figure}

\begin{figure}[t!]
    \centering    \includegraphics[width=0.85\columnwidth]{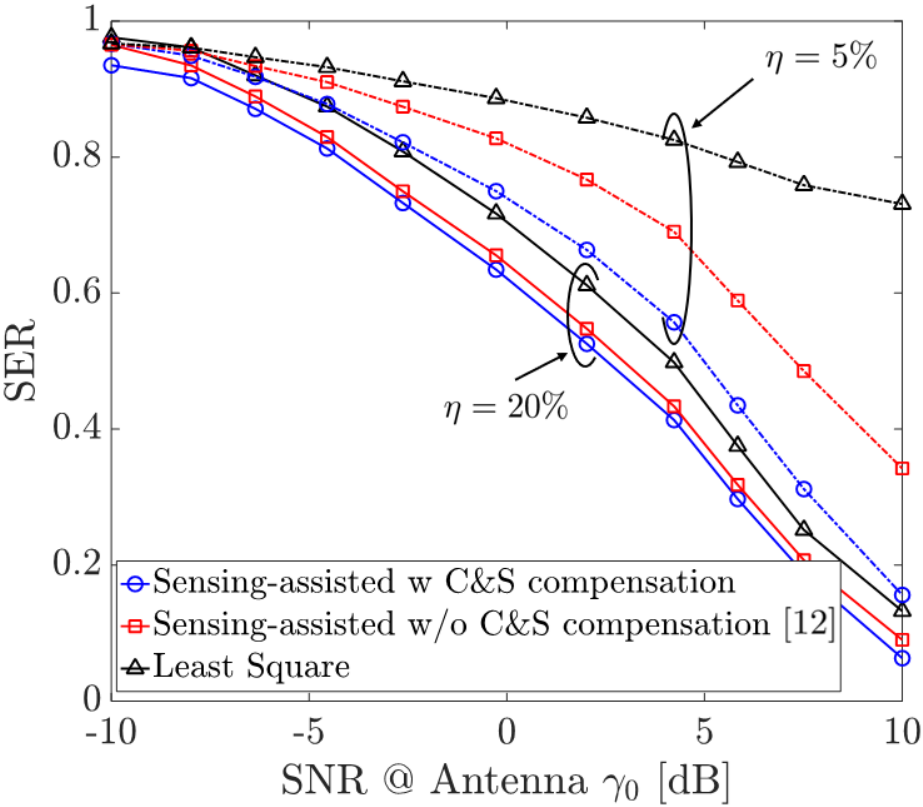}
    \caption{SER  vs Communication SNR at the antenna.}
    \label{fig:se}
    \vspace{-0.5cm}
\end{figure}

The proposed solution is compared to the conventional LS and the sensing-aided algorithm of~\cite{9805471}.
The results in Fig. \ref{fig:mse} and \ref{fig:se} reveal the limitations of the classical LS method when pilot resources are insufficient. In contrast, both the proposed method and~\cite{9805471} exhibit notable performance improvements, even with low pilot density. This highlights the efficacy of integrating sensing information, facilitating reliable channel estimation with minimal overhead.
Remarkably, the proposed algorithm's ability to compensate for C\&S mismatch leads to a substantial enhancement in MSE in Fig.~\ref{fig:mse}, resulting in a reduction of approximately $3$ dB and 5 dB for $\eta = 5, 20\%$, respectively, at $\gamma_0 = 10$ dB. A comparable improvement is noticeable in Fig.~\ref{fig:se}, where the proposed algorithm reduces the SER by half for $\eta = 5\%$ and $\gamma_0 = 10$ dB. The proposed method attains similar SER performance with respect to the state-of-the-art solutions while requiring 4$\times$  less overhead.

\section{Conclusion}\label{sect:conclusion}
This paper introduces a new approach to channel estimation in indoor mmWave scenarios exploiting the novel ISAC paradigm. The method considers and corrects for differences between the communication and sensing channel modes. Extensive numerical simulations reveal that the proposed method provides better MSE and SER performance than both the conventional LS approach and the sensing-assisted state-of-the-art method, which do not account for these differences. In particular, in the considered setup, the proposed method is able to achieve similar levels of MSE and SER performance with $4 \times$ less overhead. 
\section{Acknowledgment}
This work is supported by ARO grants, W911NF2110071, W911NF2320103, NSF grants, ECCS-2033433, and ECCS-2320568.

\bibliographystyle{IEEEtran}
\bibliography{biblio}

\end{document}